\documentclass[a4paper,11pt]{article}
\usepackage{jheppub}
\usepackage{lineno}
\usepackage{multirow}
\usepackage{comment}
\usepackage{graphicx}
\usepackage{placeins}

\arxivnumber{2502.12169} 

\title{\boldmath Antimatter Annihilation Vertex Reconstruction with Deep Learning for ALPHA-g Radial Time Projection Chamber}

\author{A. Ferreira,}
\author{M. Singh,}
\author{Y. Saito,}
\author[1]{A. Capra\note{Corresponding author.},}
\author{I. Carli,}
\author{D. Duque Quiceno,}
\author{W. Fedorko,}
\author{M. C. Fujiwara,}
\author{M. Li,}
\author{L. Martin,}
\author{G. Smith,}
\author{A. Xu,}

\affiliation{TRIUMF,\\
4004 Wesbrook Mall, Vancouver, BC V6T 2A3, Canada}

\emailAdd{acapra@triumf.ca}

\abstract{
The ALPHA-g experiment at CERN aims to precisely measure the terrestrial gravitational acceleration of antihydrogen atoms. A radial Time Projection Chamber (rTPC), that surrounds the ALPHA-g magnetic trap, is employed to determine the annihilation location, called the vertex. The standard approach requires identifying the trajectories of the ionizing particles in the rTPC from the location of their interaction in the gas (spacepoints), and inferring the vertex positions by finding the point where those trajectories (helices) pass closest to one another. In this work, we present a novel approach to vertex reconstruction using an ensemble of models based on the PointNet deep learning architecture. The newly developed model, PointNet Ensemble for Annihilation Reconstruction (PEAR), directly learns the relation between the location of the vertices and the rTPC spacepoints, thus eliminating the need to identify and fit the particle tracks. PEAR shows strong performance in reconstructing vertical vertex positions from simulated data, that is superior to the standard approach for all metrics considered. Furthermore, the deep learning approach can reconstruct the vertical vertex position when the standard approach fails.
}

\begin{document}
\maketitle
\flushbottom

\section{\label{sec:intro}Introduction}
The ALPHA-g experiment represents an extension of the ALPHA antihydrogen ($\overline{\mathrm{H}}$) program at CERN. This novel apparatus is designed to measure the gravitational acceleration of $\overline{\mathrm{H}}$ to a precision of 1\%. The determination of the gravitational properties of antimatter has always been of great interest, both as a confirmation of the validity of the fundamental principles of the General Theory of Relativity, and as an experimental milestone. Recently, the ALPHA-g experiment produced the first measurement of the terrestrial gravitational acceleration of the $\overline{\mathrm{H}}$ atom \cite{anderson2023observation}.

In this work, we introduce the \textbf{PointNet Ensemble for Annihilation Reconstruction} (\textbf{PEAR}), our first attempt to directly predict the vertical position of the $\overline{\mathrm{H}}$ annihilation for the ALPHA-g experiment using deep learning. This bypasses the challenging intermediate steps required by the current conventional tracking method, such as track identification and best-fitting procedures. This approach potentially allows us to utilize all the pieces of information available from the tracking detector, not limited to the tracks of the particles of typical interest, namely charged pions ($\pi^\pm$). Using simulated data, we show that PEAR provides superior performance to the conventional method in determining the vertical position of the $\overline{\mathrm{H}}$ annihilation. 

The method described in this manuscript relies on simulated data. The conventional method, also developed based on simulated data, was successfully deployed in \cite{anderson2023observation}, certifying an acceptable level of agreement between simulated and experimental datasets. Nevertheless, the current deep learning approach will also be tested with experimental data in future work. The goal of the present work is to lay the groundwork for integrating deep learning into the ALPHA-g experiment’s analysis pipeline.

Machine learning techniques have been successfully applied to other nuclear/particle physics experiments. In low-energy nuclear physics experiments, convolutional neural networks (CNNs) have been used to classify the tracks of nuclear decays detected by the Active-Target Time Projection Chamber (AT-TPC) \cite{Kuchera2019}. CNNs have also been employed in the context of the ASACUSA antihydrogen experiment to distinguish between antiproton and antihydrogen annihilation events \cite{sadowski2017}.

This paper is organized as follows. In the following Section~\ref{subsec:alpha-g_setup}, a brief introduction of the ALPHA-g experiment is given. We focus mainly on the radial Time Projection Chamber (rTPC) \cite{capra2017design}, which is the primary $\overline{\mathrm{H}}$ annihilation detector. The source of the training and evaluation data, namely the Monte Carlo simulation of the rTPC, is described in Sec.~\ref{subsec:MC}. Since the main interest of the current study is the improvement of \textit{$\overline{\mathrm{H}}$} annihilation vertex reconstruction techniques, a concise description of the standard vertex reconstruction is provided in Sec.~\ref{subsec:helix}. This includes the reconstruction of the position of individual ionization clusters, referred to as \textit{spacepoints}, utilized both in the standard vertex reconstruction and in the deep learning method. In Sec.~\ref{sec:method}, we describe PEAR -- a novel method to determine the position of annihilation based on deep learning. The deep learning architecture, based on \textit{PointNet} \cite{Qi2017}, is presented in Sec.~\ref{subsec:deeplearning}. Details of the data preprocessing procedure are given in Sec.~\ref{subsec:preproc}. The description of the models and training procedures is provided in Sec.~\ref{subsec:training-description}. The results and discussion of the performance of the deep learning models are provided in Sec.~\ref{sec:results}. We provide concluding remarks in Sec.~\ref{sec:conclusion}, as well as future directions.

\subsection{The ALPHA-g Experimental Setup\label{subsec:alpha-g_setup}}
For the purpose of the present discussion, we only provide a brief description on how the $\overline{\mathrm{H}}$ atoms are prepared in the experiment. For more details on the methods to synthesize, magnetically confine, and accumulate $\overline{\mathrm{H}}$ atoms, see \cite{amole2014alpha, ahmadi2017antihydrogen}. Antiprotons ($\overline{p}$) and positrons ($e^+$) are injected into the ALPHA-g Penning trap from the lower end of the cryogenic apparatus, inserted inside the rTPC. $\overline{\mathrm{H}}$ production occurs near the centre of the ALPHA-g Ioffe trap, which is generated by a set of superconducting magnets. The antiatoms, owing to their magnetic dipole moment, are confined near the minimum of the magnetic field gradient.  The experiment described in \cite{anderson2023observation} was performed by slowly removing the confining field and counting the annihilation events as a function of the vertical coordinate. The release of $\overline{\mathrm{H}}$ was repeated multiple times, while changing the relative strength of the trapping coils. 

The rTPC is the position-sensitive detector of the ALPHA-g experimental setup, which is capable of imaging the tracks of the products of the $\overline{\mathrm{H}}$ annihilation, thus recovering the information on the location of the annihilation \cite{capra2017design}. Typically, the antiproton annihilation with a nucleon (proton or neutron) of the electrodes of the Penning trap results in the release of a number of neutral $\pi^{0}$ and charged $\pi^{\pm}$ mesons, the latter being the particle of interest for the reconstruction of the annihilation position.

The detector is placed vertically, with its axis parallel to the gravitational field of the Earth. The cylindrical chamber is 2.3~m long and is filled with a gas consisting of a mixture of argon and carbon dioxide at atmospheric pressure. The cryogenic $\overline{\mathrm{H}}$ magnetic trap is housed inside the cylindrical volume, delimited by the inner cathode.
The charged particles, mostly $\pi^{\pm}$, produced in $\overline{\mathrm{H}}$ annihilation ionize the gas in the rTPC. The electrons, freed in the ionization process, drift radially towards the anode wires (located near the outer edge of the active volume), under the influence of a radial electric field and an axial magnetic field of 1~T. The avalanche signal induced on the anode wires is also induced on the segmented electrodes, called \textit{pads}, on the outer surface (facing inwards) of the rTPC.

\subsection{\label{subsec:MC}Monte Carlo Simulation of Antihydrogen Annihilation}
The Monte Carlo simulation of the response of the rTPC is implemented using the GEANT4 toolkit~\cite{AGOSTINELLI2003250,1610988,ALLISON2016186}. Annihilation products are generated at the wall of the apparatus based on the branching ratios~\cite{bendiscioli1994antinucleon} of the proton-antiproton interactions, and the simulation of their trajectories assumes a 1~T magnetic field. A detailed CAD model (including Penning trap electrodes, superconducting coils, vacuum chambers, etc.) is provided to GEANT4 to simulate the passage of particles through the materials~\cite{poole2012acad}. The GEANT4 implementation of the Photo Absorption Ionization (PAI) model~\cite{APOSTOLAKIS2000597} is employed to simulate the interaction of the charged particles in the drift volume of the detector (gas medium). The ionization drift towards the anode wires is taken into account using a pre-computed table, calculated with the Garfield++ package~\cite{schindlergarfield++}. The signals on the wires and on the pads are generated due to the movement of the charges in the rTPC, following the charge multiplication process (avalanche) in the vicinity of an anode wire. A signal template, different for wires and pads, is computed using the Ramo theorem \cite{sauli2015gaseous}, which permits the derivation of the signal induced on electrodes by a single charge. The functional form of the signals takes into account the shaping electronics, and their amplitude is determined using real data.

The simulation produces an output file in the same format as the real data (including data acquisition binary protocols, waveform digitization, etc.). This allows us to validate the simulation using the same analysis tools that are used to process real $\overline{\mathrm{H}}$ annihilation events.

\subsection{\label{subsec:helix}Standard Reconstruction of Antihydrogen Annihilation Vertex}

The current standard method of the $\overline{\mathrm{H}}$ vertex reconstruction proceeds in two stages. The first stage reconstructs the three-dimensional position of individual ionization, otherwise known as spacepoints, caused by the interaction of charged particles with the rTPC gas. The signal templates deployed for the Monte Carlo simulation are also used to determine the arrival times of individual electron clusters on the anode wires and on the pads. Since several pads are involved in each avalanche, positional information is extracted by combining neighbouring pad hits, and a unique value for the $z$ coordinate is obtained for each electron avalanche. Each wire signal is matched to the nearest groups of pads by their azimuthal position $\phi$ and the drift time. Finally, the inverse pre-computed table, used in the simulation for electron drift, is used to perform the \textit{time projection}, thus converting the drift time into the radial coordinate $r$. Moreover, since the rTPC is in a 1~T field, it is necessary to correct for the charge displacement in $\phi$, due to the Lorentz force acting in the drift plane, which is orthogonal to the magnetic field. The triplet of numbers $(r,\phi,z)$ uniquely determines a spacepoint, and it is saved in Cartesian coordinates. This first stage of analysis is used in both the conventional vertex reconstruction and the deep learning method presented in this work.

The second stage of the reconstruction begins by grouping spacepoints together into tracks. Each track is then fit to a helix function using the least squares method. Finally, the vertex location is found from the helix analytical functions. The pair of helices that pass closest to each other determines the initial location of the vertex. This can be improved by adding additional helices, minimizing their relative distance again. This second stage of analysis, which reconstructs vertices by fitting helix functions, will be referred to in the rest of the manuscript as the \textit{standard} method, or the \textit{Helix Fit} method.

The standard method is thus based on realistic assumptions on the kinematics of the particles under the influence of the electromagnetic fields. Once the individual $\pi^{\pm}$ tracks in the rTPC signals are clearly identified, this method provides an intuitive way to reconstruct the annihilation vertices. However, identification and clustering of tracks is generally challenging due to electronic noise, multiple scattering, $e^+e^-$ pair-production following neutral pions ($\pi^{0}$) decays, and the variable number of ($\pi^{\pm}$) particles as a result of $\overline{\mathrm{H}}$ annihilation. It is worth noting that the sign of the charged pion is not relevant for the reconstruction of the annihilation vertex, and no attempt is made to determine whether the antiproton has annihilated with a proton or a neutron. If a sufficient number of $\pi^{\pm}$ tracks cannot be identified (i.e., at least two), this method cannot reconstruct the positions of the annihilation vertices. Noise, coupled to the active environment of the experimental hall, and multiple scattering may also degrade the quality of the fit, resulting in increased error in the position of the reconstructed vertices.

\section{\label{sec:method}Methods}
\subsection{Deep Learning Approach for Vertex Reconstruction\label{subsec:deeplearning}}
One of the approaches to overcoming the challenges outlined at the end of the previous Sec.~\ref{subsec:helix} is to bypass the identification of tracks and the subsequent fitting procedure, by directly predicting the $\overline{\mathrm{H}}$ annihilation vertex position. This is achieved by learning the relationship between the spacepoints in an rTPC event and the corresponding vertex position.

\begin{figure}[htbp]
    \centering
    \includegraphics[width=\textwidth]{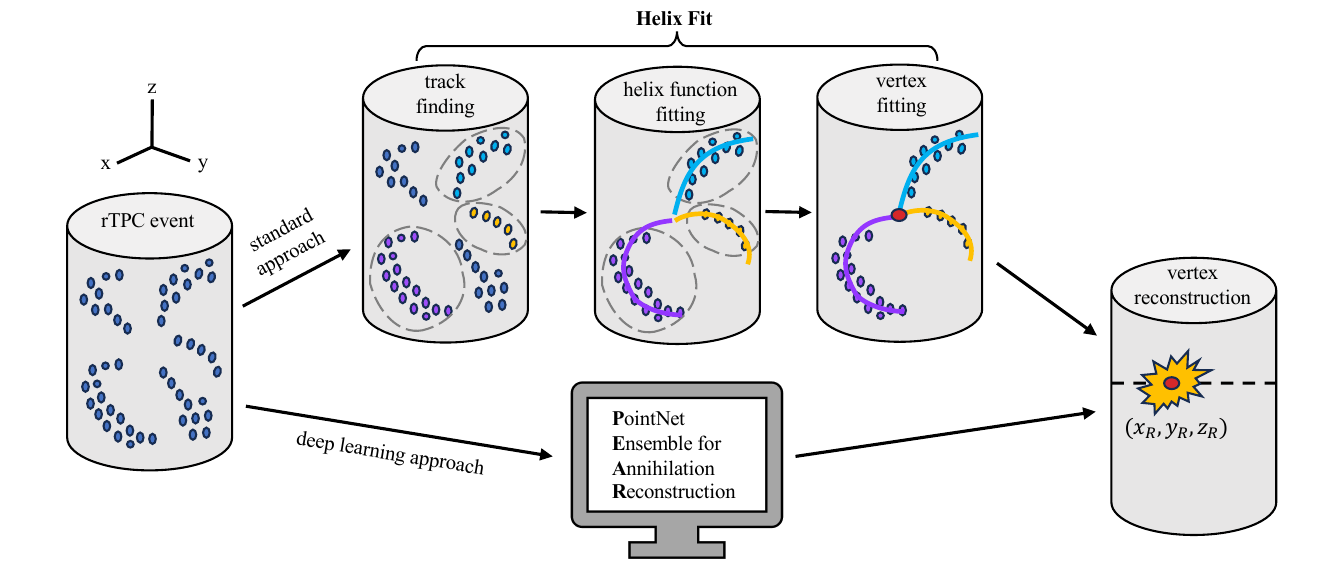}
    \caption{\label{fig:concept}Conceptual schematics of the vertex reconstruction approach using our deep learning model (bottom), in contrast to the standard method that requires identification of particle tracks and fitting helix functions (top).}
\end{figure}

Deep learning models are suitable for such tasks. This will not only eliminate the necessity of identifying the $\pi^{\pm}$ tracks, but could also allow one to utilize the information from all the spacepoints that are generated in the annihilation process. The conceptual schematics of this approach, along with the procedure of the standard method, is shown in Fig.~\ref{fig:concept}.

Our deep learning architecture is based on \textit{PointNet} \cite{Qi2017}, which was originally developed to classify and segment the shape of a set of data points in a three-dimensional (3D) space. PointNet is able to directly take 3D points as input, which can avoid the potential loss of information caused by projecting them onto 2D planes, as would be needed to apply a 2D convolutional neural network. Furthermore, learning directly from sparse and irregular 3D points with PointNet is significantly more efficient than representing them as 3D voxel grids \cite{Qi2017} and applying a model such as a 3D convolutional neural network.

In the current development, the goal is to predict the position of $\overline{\mathrm{H}}$ annihilation vertices from the 3D input data. Since this can be framed as a regression task, we need to modify the PointNet classification network for regression tasks.
For our application, we focus on the reconstruction of the $z$-coordinate of the vertex, since it represents the position along the vertical axis of the rTPC, where the effect of gravity is measured. In this work, the $x$ and $y$ coordinates of the vertex position are not reconstructed, as they are not directly relevant for measuring the gravitational acceleration of $\overline{\mathrm{H}}$. The geometry of the rTPC and the orientation of the electric (drift) and magnetic fields are such that the reconstruction problem separates into its orthogonal components: the radial plane ($x-y$ plane), and the axial direction $z$. This implies that the $x-y$ reconstruction cannot be used directly to infer the quality for our method to determine the $z$ coordinate of the vertex.
In what follows, we will discuss the relevant features of the PointNet architecture and the modifications made to achieve our goal.

\begin{figure}[htbp]
    \centering
    \includegraphics[width=\textwidth]{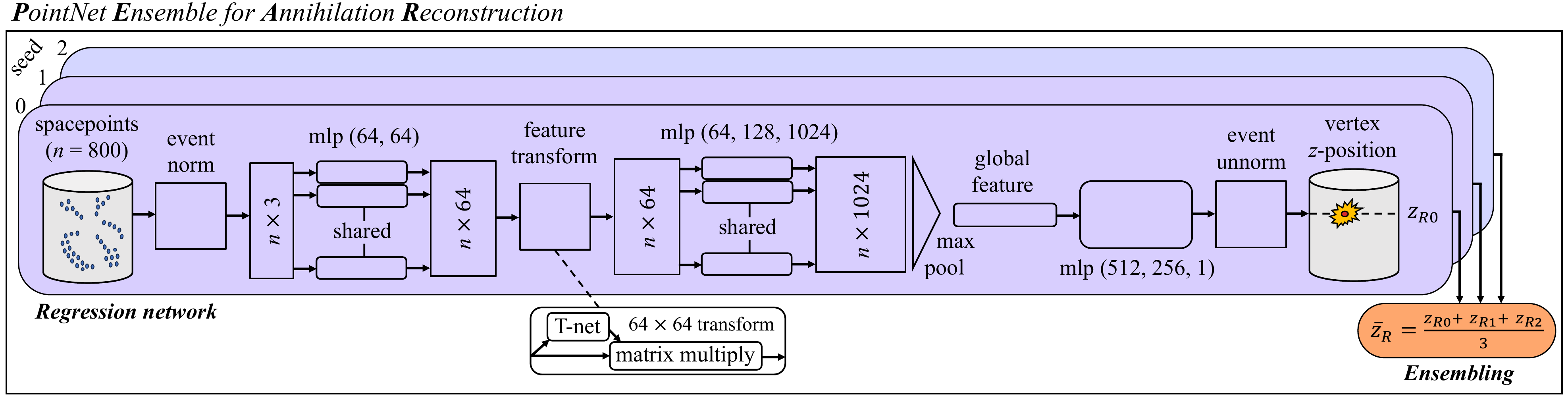}
    \caption{\label{fig:architecture} Schematics of our modified PointNet architecture for the vertex reconstruction regression task, heavily based on \cite{Qi2017}. The dimensions of hidden and output layers of the MLPs are indicated in parentheses. The architecture has been simplified by removing the initial input transform, and the final layer has been replaced with a linear layer to predict the $z$ position of the vertex. Unlike the original PointNet, which outputs classification scores, PEAR outputs a single prediction for each event, focused solely on $z$-vertex reconstruction. Three models are trained with this architecture, using three different random seeds, and their outputs are averaged to get the final $z$-vertex prediction.} 
\end{figure}

Figure~\ref{fig:architecture} shows the architecture of our model based on PointNet. The design of PointNet aims to extract higher-dimensional representations of the input point cloud (spacepoints) on a per-point basis, and then to aggregate the information from all the points to make a prediction on the quantity of interest. 

Each point is processed with two Multilayer Perceptrons (MLPs), interleaved with a so-called \textit{T-Net}. The T-Net outputs a square matrix which can be thought of as a linear geometric transformation of each point in the point cloud. An MLP is an archetypical artificial neural network composed of multiple layers of neurons, including an input layer, one or more hidden layers, and an output layer, where each neuron applies a weighted sum, followed by a non-linear function, called the  \textit{activation function}. The first MLP has three input nodes corresponding to the three spatial coordinates of a spacepoint. In this architecture the first two MLPs produce \textit{features} - that is high dimensional representations of the input data.

The original PointNet implementation applies a T-Net on the input data, making it possible to account for the invariance of the data to geometric transformations such as translation and rotation. In this work, we perform the matrix transform only on the extracted features, not on the input data. The input data transform is replaced by the data preprocessing discussed in Section.~\ref{subsec:preproc}.

Following point-wise MLP and T-Net processing, the points are aggregated using so-called max-pooling operation, where, for each feature, the maximum value of that feature from the entire point cloud is selected. The aggregation of the points with the max-pooling layer also makes the model invariant to the ordering of the input data points. Although the rTPC spacepoints could be ordered by the drift time or amplitudes of the anode wire (wire amplitude) signals, in this work, we only use the position information of spacepoints, except when selecting spacepoints or padding events (see Sec.~\ref{subsec:preproc}). The final MLP then takes the global feature vector and produces a scalar output. For more details of the architecture, see Refs.~\cite{Qi2017, pointnetpp}.

Further improvements in model performance can be made by averaging the predictions of separately trained models to form an \textit{ensemble}. Ensemble learning is an approach for combining the predictions of multiple models to achieve better performance than any individual model alone \cite{Breiman1996BaggingP, ensemble_survey_1, ensemble_survey_2}. Ensemble learning can lead to a reduction in variance, an improvement in precision, and a more robust prediction.
Fig.~\ref{fig:architecture} presents a schematic summary of PEAR, the PointNet Ensemble for Annihilation Reconstruction, an ensemble of three copies of the same PointNet-like regression network architecture, initialized with different random seeds. 
In the current work, we note that the use of ensembles improved the resolution and robustness of the model predictions for unseen data.

\subsection{\label{subsec:preproc}Data Preprocessing}
A dataset of 2.7 million $\overline{\mathrm{H}}$ annihilation events was generated using the Monte Carlo simulation described in Sec.~\ref{subsec:MC}. The dataset consists of the true $\overline{\mathrm{H}}$ annihilation vertices, and the corresponding coordinates of the spacepoints from the simulated rTPC signal, as well as the anode wire amplitude. The distribution of the $z$ positions of the simulated vertices is approximately uniform across the vertical length of the rTPC. In total, 2.2 million events were used as training data, 260 thousand events were used as validation data, and 290 thousand events were used as test data. The test dataset is completely separate from the other datasets and was only used once, to perform the final evaluations for reporting the current results. We divided our validation set into two equal parts, Validation set A ($\mathrm{Val}_{A}$) and Validation set B ($\mathrm{Val}_{B}$), while ensuring that both subsets had a similar distribution of events. During training, we used $\mathrm{Val}_{A}$ to compute validation metrics. After training, we cross-check these metrics by calculating them again with $\mathrm{Val}_{B}$ to confirm the performance of the model on unseen data. The test dataset was left untouched until the final performance evaluation.

The data must be preprocessed before being fed into PEAR. First, a fixed number of spacepoints in each event must be determined since PointNet requires the input shape of all the events to be consistent. In choosing the number of spacepoints in a given event, it is important to note that, while a larger number of spacepoints would preserve more complete information, it would also slow down the training iterations due to the larger data size. 
Taking this into account, we choose the number of spacepoints per event to be 800, since around 95\% of the events in the training data set contain 800 or fewer spacepoints. If the number of spacepoints in a given event is larger than 800, the points with the smallest anode wire amplitude are discarded, and if the number of spacepoints is smaller than 800, points are randomly duplicated until exactly 800 points are obtained. Padding with duplication is indeed the approach adopted in the original PointNet implementation \cite{Qi2017}. This is because the duplicated spacepoints will not affect the outcome due to the max-pooling layer. It is preferable to ``padding with zeros'', since the point $(0,0,0)$ lies outside the active volume of the rTPC.

Among several normalization schemes we explored, averaging the $z$ coordinate of the 800 input spacepoints in a given event and subtracting the mean from both the $z$ coordinates of the spacepoints and the true vertex led to a significant improvement in performance. The prediction of the $z$ coordinate of the vertex can then be obtained by adding the mean $z$ of the spacepoints to the output of PEAR. Focusing on the position of the spacepoints and vertices relative to the mean $z$ makes the data invariant to the translation in the $z$ direction, resulting in a reduction in bias in the prediction of the $z$ coordinate of the vertices. The $x$ and $y$ coordinates are left unchanged, as they are not expected to be invariant to translations. 

\subsection{\label{subsec:training-description}Training and Construction of an Ensemble}
The model and training iteration were implemented with PyTorch \cite{NEURIPS2019_bdbca288}, based on an existing PointNet implementation by Ref.~\cite{Xu_pointnet_2019}. We also used various open-source Python libraries \cite{numpy, scipy, matplotlib, h5py, iminuit}.
Training was carried out with the Adam optimizer, using the default values of the parameters implemented in PyTorch: initial learning rate of 0.001, coefficients to calculate the running averages of the gradients and their squares $\beta_1 = 0.9$ and $\beta_2=0.99$. Based on our manual hyperparameter search, the best results were obtained when the batch size was 512 and the learning rate was multiplied by 0.8 every 100 epochs.

The Huber loss function \cite{robust_regression}, defined by
\begin{equation*}
    \label{eq:huber_loss}
    L_\delta(\Delta z) =
    \begin{cases} 
        \frac{1}{2} (\Delta z)^2 & \text{for } |\Delta z| \leq \delta \\
        \delta (|\Delta z| - \frac{1}{2} \delta) & \text{for } |\Delta z| > \delta,
    \end{cases}
\end{equation*}
is used as the objective function to minimize the difference, or \textit{residual}:
\begin{equation}
    \label{eq:res}
    \Delta z = z_{\mathrm{Reconstructed}} - z_{\mathrm{Simulated}},
\end{equation}    
between the simulated $z_{\mathrm{Simulated}}$ and predicted $z_{\mathrm{Reconstructed}}$ vertex coordinate during training. This loss combines the advantages of the Mean Absolute Error (MAE) and the Mean Squared Error (MSE) loss functions, which makes the training of the model less sensitive to the outliers in data than the MSE, while still providing smoothness near zero \cite{pytorch_huber_loss}. After empirically confirming that Huber loss outperforms MAE and MSE, we focused on the optimization of the hyperparameter $\delta$. As a result of conducting a manual hyperparameter search, we determined that setting $\delta=1$ yielded the best performance. This setting means that for residuals with an absolute value smaller than one, the loss behaves quadratically, whereas for larger residuals, it switches to a linear response. While the minimization of the loss was performed using the training dataset, we also tracked the loss on the $\mathrm{Val}_{A}$ set, a portion of the data that is unseen to the model, to ensure that the model generalizes sufficiently well. The training was performed using an NVIDIA A100 GPU with 100~GB memory, available on the Narval cluster managed by the Digital Research Alliance of Canada. The duration of training was approximately 70 hours. 

Additionally, during training, we tracked performance across different metrics in addition to the Huber loss. Of particular relevance is the deviation of the mean of the distribution of the residuals ($\Delta\mathrm{z}$) with respect to zero, which we refer to as the $z$-bias. An accurate prediction of the $z$ coordinate of the annihilation vertex has a null bias. Since the $z$-bias may depend on the $z$ position along the vertical axis of the rTPC, we further devised our $z$-bias measure to take the dependence into account, that we name \textbf{Absolute Residual Average} (\textbf{ARA}). To calculate the ARA, we first divide the detector along the $z$-axis into slices. Each slice is defined as a 100~mm segment of the rTPC, which allows us to examine the performance of the model at different $z$ locations throughout the entire 2304~mm vertical length of the detector. The number of 100~mm slices is 24 in total, however, we exclude the first four and last four to avoid edge effects (loss of geometric acceptance), which now makes the total number of slices $N_{\mathrm{slices}} = 16$. For each slice $i$, we compute the residuals as defined in Eq.~(\ref{eq:res}) for each event to obtain the distribution of $\Delta\mathrm{z}_i$. The absolute value of the mean of the residuals in each slice $|\overline{\Delta \mathrm{z}_i}|$ is then averaged over the slices to obtain the ARA:
\begin{equation}
    \label{eq:overall_bias}
    \mathrm{ARA} = \frac{1}{N_{\mathrm{slices}}}\sum_{i}^{N_{\mathrm{slices}}} |\overline{\Delta \mathrm{z}_i}|.
\end{equation}

We initialized our deep learning model with three different random seeds: 0, 1, and 2, and trained them separately. The optimized models initialized with each of the random seeds form an ensemble, i.e., PEAR. Training was carried out for at least 400 epochs, until the overall loss of the training set converges to stable values or stops consistently improving over 200 epochs. The number of training epochs did not exceed 1000. Whenever the model exhibited values of ARA lower than the standard (Helix Fit) method, we saved the weights of the model as a checkpoint. Of the saved checkpoints, the one with the smallest ARA value in the $\mathrm{Val}_{A}$ dataset was chosen to form an ensemble. The results presented in the following section were obtained with PEAR constructed as discussed above.

\section{\label{sec:results}Results and Discussion}
In this section, we evaluate the performance of PEAR using the test dataset, which contains 290 thousand events. We show that it achieves results significantly better than, or at least similar to, the standard (Helix Fit) method on all metrics considered. Note that we are primarily concerned about performance in the region from -800~mm to 800~mm, as this is roughly the region where the physics experiment is carried out. 

\subsection{Average Performance\label{subsec:avg_performance}}
\begin{figure}[htbp]
    \centering
    \includegraphics[width=\textwidth]{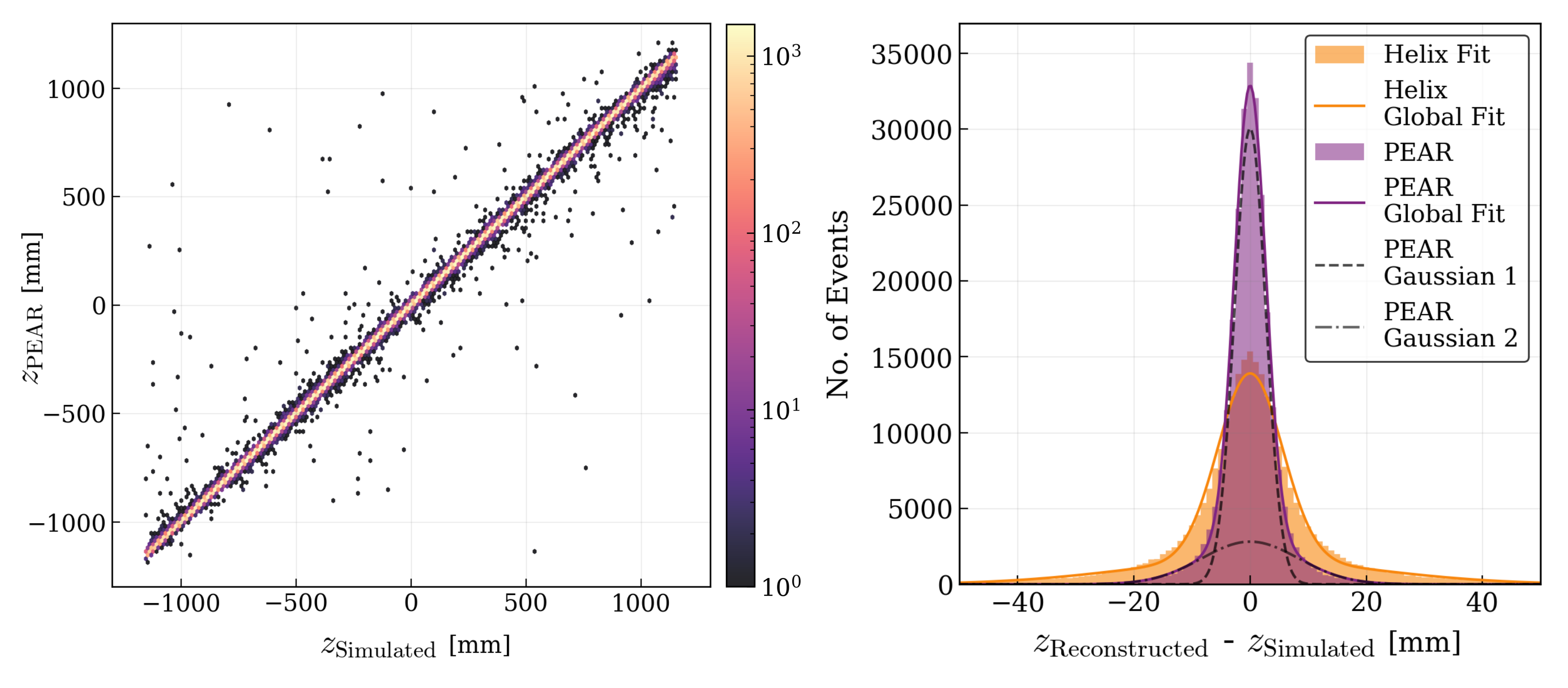}
    \caption{\label{fig:res_dist} 2D histogram of predicted $z$ coordinate of the vertices by PEAR versus simulated (true) $z$ coordinate (left). Histogram of residuals for PEAR and Helix Fit with Gaussian fits (right). These plots use the test dataset.}
\end{figure}

Fig.~\ref{fig:res_dist} (left panel) shows the $z$ coordinate of the vertices predicted by PEAR on the vertical axis and the simulated (true) $z$ coordinate on the horizontal axis, across the whole rTPC length. It can be seen that the vast majority of the points lie on the $y=x$ line throughout the length of the detector. Fig.~\ref{fig:res_dist} (right panel) shows the distribution of the residual (Eq.~\ref{eq:res}), i.e., the deviation of the prediction from the true $z$ coordinate of the vertices. The narrower distribution of PEAR predictions compared to the standard method indicates that PEAR provides a more precise reconstruction of the $z$ position of the vertices. 

We model and fit the distribution of the residuals as a sum of two Gaussians (bi-Gaussian) that share a common mean $\mu$, along with a constant offset term. Table~\ref{tab:res_results} shows the results of the bi-Gaussian best fits for PEAR and the standard method. The standard deviations of each of the Gaussians ($\sigma_1$ and $\sigma_2$), the overall standard deviation ($\sigma$), and the Full Width at Half-Maximum (FWHM) are all less than half of those of the standard method. This quantitatively indicates a significant improvement in the precision of the reconstruction of the $z$ coordinate of the annihilation vertices. The integrals shown in the table indicate the fractions of the events corresponding to each of the Gaussian components. The narrower Gaussian in PEAR occupies a larger fraction than the corresponding Gaussian does in the standard method, which indicates that a larger fraction of the events can be reconstructed with a better precision with PEAR. The ARA, which indicates the average $z$-bias across different portions of the detector, is also smaller for PEAR, suggesting that the performance is consistently superior throughout the detector length. The performance in different parts of the detector is discussed in more detail below. The mean $\mu$ of the PEAR is compatible with zero within two sigma of the statistical error, as expected for an unbiased prediction of the annihilation vertex position. However, in the test dataset analyzed above, 0.3\% of events are have a residual over 50~mm, with the largest outliers reaching over 2000~mm.

\begin{table}[h!]
    \centering
    \begin{tabular}{|l|l|l|l|}
        \hline & Metric & \textbf{PEAR} & \textbf{Helix Fit} \\ 
        \hline
        \multirow{2}{*}{Gaussian 1} & $\sigma_1$ [mm] & $2.65$ $\pm$ $0.01$ & 5.52 $\pm$ 0.02\\ 
        \cline{2-4} & Integral & 76.9\%& 66.3\%\\ 
        \hline
        \multirow{2}{*}{Gaussian 2} & $\sigma_2$ [mm] & $8.17$ $\pm$ $0.07$& $21.17$ $\pm$ $0.13$\\ 
        \cline{2-4} & Integral & 22.5\%& 32.9\%\\ 
        \hline
        \multirow{4}{*}{Overall}
        & $\sigma$ [mm] & $3.50 \pm 0.02$ & $8.85 \pm 0.04$ \\ \cline{2-4} 
        & $\mu$ [mm] & -0.02 $\pm $0.01& -0.02 $\pm$ 0.02\\
        \cline{2-4}
        & FWHM [mm] & $6.62 \pm 0.02$ & $14.12 \pm 0.06$\\ \cline{2-4} 
        & ARA [mm] & $0.04 \pm 0.01$ & $0.06 \pm 0.02$\\  \hline
    \end{tabular}
    \caption{\label{tab:res_results} Comparison of predictive performance between PEAR and the standard method (Helix Fit). Note the ARA calculation of Eq.~(\ref{eq:overall_bias}) is at 100~mm slices and only includes the volume from -800~mm to 800~mm. Additionally, there is a constant component to the fit accounting for less than 1\% of the integral.}
\end{table}

\subsection{Position-dependent Performance\label{subsec:pos_performance}}
\begin{figure}[htbp]
    \centering
    \includegraphics[width=0.84\textwidth]{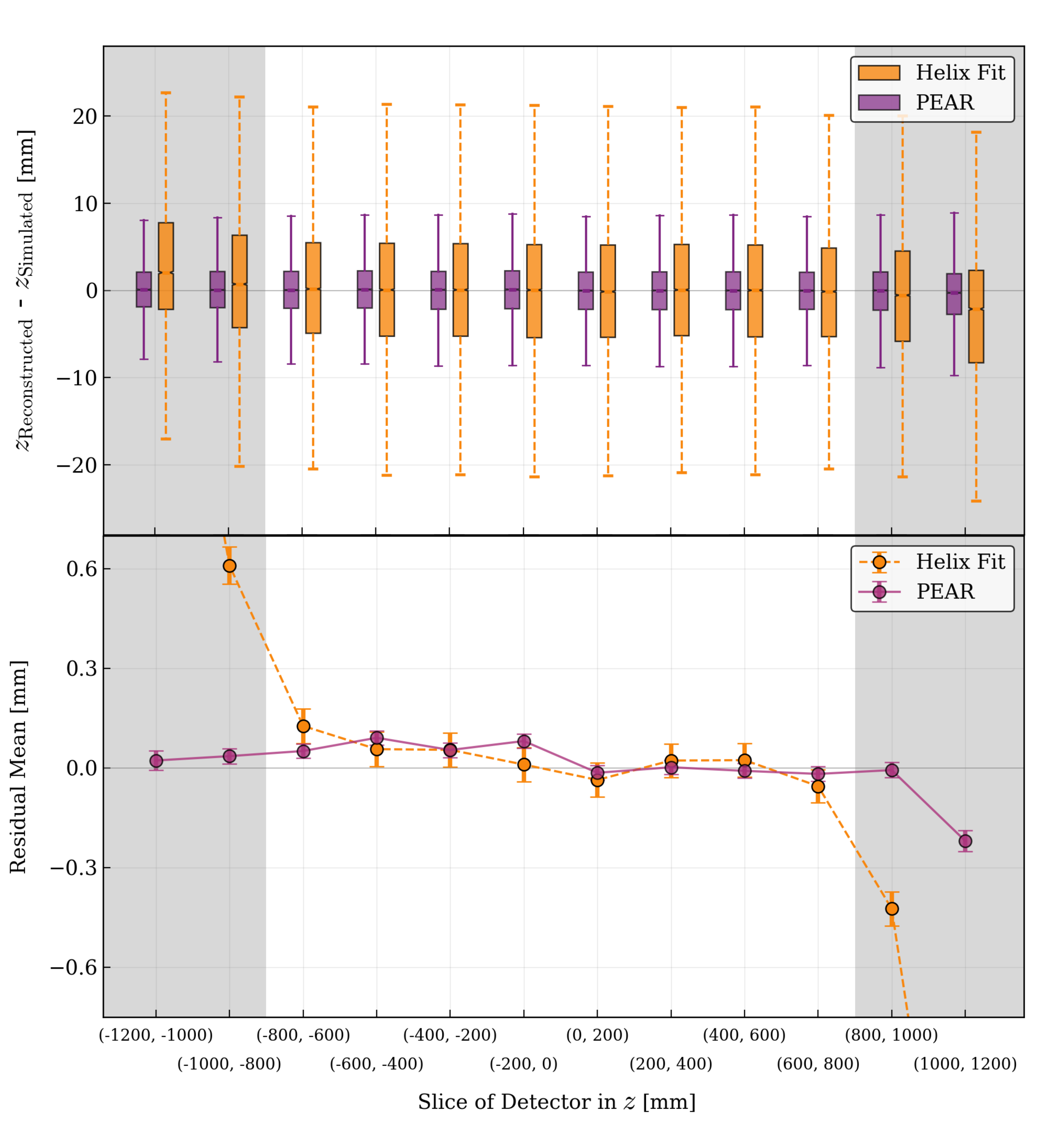}
    \caption{\label{figs:box_and_bias}Box plot of residuals from PEAR and Helix Fit predictions (top). For this study, the line within the box denotes the median, the box extends from the lower to upper quartile, otherwise known as the interquartile range (IQR), and the whiskers extend to the furthest residuals within 1.5 times the IQR on either side of the distribution. Gaussian mean (used in ARA Eq.~(\ref{eq:overall_bias}) to probe $z$-basis) (bottom). Each for 200~mm slices of the detector.}
\end{figure}

To investigate the performance of PEAR in different sections of the detector, we divide the rTPC along the $z$-axis (axis along the length) into 200~mm wide slices. Fig.~\ref{figs:box_and_bias} (top panel) shows a box plot obtained from the distribution of the $z$ residuals in each part of the detector. Each box contains data from the 25th to the 75th percentile, which is referred to as the interquartile range (IQR), and the lower (upper) end of the whisker represents the lowest (highest) data point within the 1.5 times the IQR from the lower (upper) end of the box. The plot shows that the distributions of the residuals with PEAR are consistently narrower than those with the standard method (Helix Fit) throughout the detector length.

Fig.~\ref{figs:box_and_bias} (bottom panel) shows the mean of the bi-Gaussian distribution along with its standard deviation, obtained in the same manner as discussed above, for each detector section. In the region of the detector between -800~mm and 800~mm, where the experiment is conducted and indicated by the white background in the figure, the deviation of the mean of the residuals is less than 0.11~mm for PEAR, and less than 0.18~mm for the standard method (Helix Fit). The calculated values of ARA (Eq.~\ref{eq:overall_bias}) are smaller than 0.1~mm for both methods. Outside the region of interest of the detector (gray background in the figure), although it is not used in the $\overline{\mathrm{H}}$ gravity experiment, PEAR demonstrates superior performance, which further validates the robustness and reliability of the method.

\subsection{Vertex Reconstruction in Complex Events\label{subsec:complexevents}}
One of the advantages of this deep learning approach is that reconstruction of vertex positions does not require identification of the tracks of the charged particles $\pi^\pm$ in the rTPC signal, which is required in the standard method. As a result, PEAR is able to predict the $z$ position of the annihilation vertices even when the standard (Helix Fit) method fails due to fewer than two $\pi^\pm$ tracks being identified. To demonstrate this, we selected seven thousand events from the test dataset where the Helix Fit method fails. These events are usually excluded from comparisons between PEAR and Helix Fit. Hereafter, we refer to these events as ``complex events''. These complex events may include noise and multiple scattering of the annihilation products, as well as incomplete tracks at the edges of the detector volume.

\begin{figure}[htbp]
    \centering
    \includegraphics[width=0.84\textwidth]{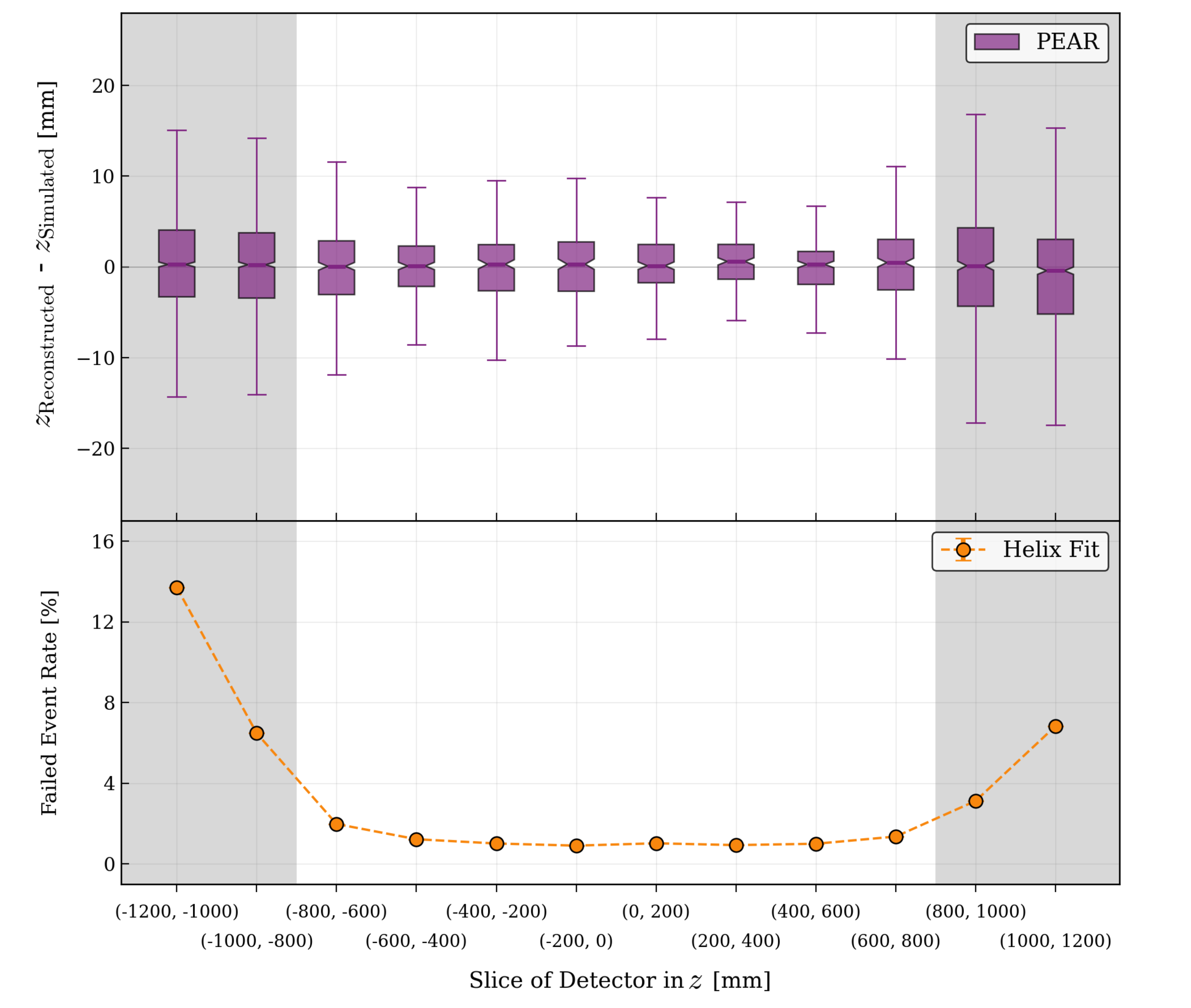}
    \caption{\label{fig:failed_samples} Analysis of seven thousand events Helix Fit is unable to provide predictions for (previously excluded). Box plot (see Figure~\ref{figs:box_and_bias} caption for general box plot description) of PEAR performance (top) and percentage of total events in each slice of the test set Helix Fit failed to predict (bottom), note that error bars on the percentages aren't visible due to their small magnitude.}
\end{figure}

Fig.~\ref{fig:failed_samples} (top panel) shows a box plot of the residuals for the complex events predicted by PEAR in each section of the detector. It demonstrates that, when compared to Figure~\ref{figs:box_and_bias} (top panel), PEAR achieves comparable performance in predicting vertex $z$ positions within the experimental detector range of -800 mm to 800 mm (depicted with a white background in the figure).  Fig.~\ref{fig:failed_samples} (bottom) shows the fraction of the events in the test dataset where the Helix Fit method fails. At the lower end of the rTPC (gray-shaded area on the left hand side), the standard method fails to reconstruct the vertices in roughly 13\% of the events in the dataset. Even for such events, the performance of PEAR is able to provide valid predictions, although the performance somewhat degrades. This suggests that, with PEAR, a wider range of the detector may be used in the experiment.

\section{\label{sec:conclusion}Conclusions and Future Work}
In this work, we explored an approach to reconstructing the position of $\overline{\mathrm{H}}$ annihilation vertices in the rTPC used in the ALPHA-g experiment, based on the PointNet deep learning architecture. Simulated annihilation events were used to train and evaluate the model for the reconstruction of the $z$ coordinate of the vertices (position along the vertical axis of the rTPC), as it is the quantity directly required for the measurement of the gravitational acceleration of $\overline{\mathrm{H}}$. 

We have developed the model: PointNet Ensemble for Annihilation Reconstruction (PEAR). We demonstrated that PEAR is capable of learning the map between the spacepoints in the rTPC signal and the true (simulated) $z$ coordinate of the vertex position. The quality of the reconstructed vertex $z$ positions was compared to the standard method, which involves identifying the tracks of $\pi^\pm$, fitting the tracks with helix functions, and then deducing the vertex positions.

PEAR achieved superior overall performance compared to the standard (Helix Fit) method on the metrics we have considered, indicating that the prediction by PEAR is more accurate and precise. The performance has been shown to be consistent across different sections along the length of the rTPC. PEAR is also capable of reconstructing the $z$ positions of the vertices even when the standard method fails due to insufficient number of identified tracks of charged annihilation products.

We have begun exploring employing the variance between predictions from the individual models in the ensemble as a potential metric for filtering out high residual events. Further work is needed to fully explore this strategy and possibly enable PEAR to produce uncertainty predictions.

In recent years, more advanced model architectures have been published \cite{pointnetpp, pointtransformerv3} that have shown performance far exceeding that of PointNet on complex tasks that involve point clouds. While these newer approaches may be worth exploring in future studies, we conclude that, for the current application, the simpler PointNet architecture is capable of performing the desired tasks and outperforms the conventional method.

Extending this model to also predict the $x$ and $y$ positions of the vertex is underway and currently showing promising results. Although these values are not directly needed for gravity measurements, they are useful for additional data analysis, such as background rejection and studies of the behavior of trapped $\overline{\mathrm{H}}$. 

Further tests and developments of PEAR with real experimental data are necessary to validate performance before use in producing scientific results. However, if this performance transfers reasonably well to real data, this model will serve as a valuable tool in aiding ALPHA-g in carrying out more precise measurements of the effect of gravity on antimatter.  

\FloatBarrier

\acknowledgments
This work was supported by NSERC, NRC/TRIUMF, CFI. 
This research was enabled in part by the computational resources provided by the BC DRI Group and the Digital Research Alliance of Canada (alliancecan.ca). We thank A. Olin for supporting this work through his Digital Research Alliance of Canada allocations.
We thank P. A. Amaudruz, D. Bishop, M. Constable, P. Lu, L. Kurchaninov, K. Olchanski, F. Retiere, and B. Shaw (TRIUMF) for their work on the ALPHA-g detectors and the data acquisition system.

\newpage
\appendix 

\section{Individual Model Runs and PEAR Performance}

\setcounter{table}{0}
\renewcommand{\thetable}{A\arabic{table}}
\begin{table}[htbp]
\centering
\scalebox{0.9}{
    \begin{tabular}{|c|c|c|c|c|c|c|}  
     \hline
    Model & Dataset & Epoch & ARA [mm] & $\sigma$ [mm] & FWHM [mm] \\ 
    \hline
     \multirow{2}{*}{Helix Fit} & $\mathrm{Val}_{A}$ &  & $0.11 \pm 0.03$ &  $8.80 \pm 0.05$& $14.09 \pm 0.09$\\
      & $\mathrm{Val}_{B}$ &  & $0.10 \pm 0.03$ & $8.80 \pm 0.06$& $14.03 \pm 0.09$\\
    \hline
     \hline
     \multirow{2}{*}{Seed 0} & $\mathrm{Val}_{A}$ & \multirow{2}{*}{134} & $0.06  \pm 0.01$ &  $4.03 \pm 0.02$  & $7.65 \pm 0.04$\\
    
      & $\mathrm{Val}_{B}$ &  & $0.05   \pm 0.01$ &  $4.04 \pm 0.02$  & $7.65 \pm 0.04$ \\
     \hline
     \multirow{2}{*}{Seed 1} & $\mathrm{Val}_{A}$ & \multirow{2}{*}{263} & $0.06  \pm 0.01$ &  $3.73 \pm 0.02$ & $7.01 \pm 0.04$\\
     
      & $\mathrm{Val}_{B}$ &  & $0.04  \pm 0.01$ & $3.76 \pm 0.06$  & $7.05 \pm 0.04$ \\
     \hline
     \multirow{2}{*}{Seed 2} & $\mathrm{Val}_{A}$ & \multirow{2}{*}{260} & $0.06  \pm 0.01$ &   $3.74 \pm 0.02$ & $7.10 \pm 0.04$ \\
     
      & $\mathrm{Val}_{B}$ &  & $0.07  \pm 0.01$ & $3.76 \pm 0.02$ &  $7.15 \pm 0.04$ \\
     \hline
     \multirow{2}{*}{\textbf{PEAR}} & $\mathrm{Val}_{A}$ & \multirow{2}{*}{[134, 263, 260]} & $0.06 \pm 0.01$ & $3.51 \pm 0.02$ & $6.65 \pm 0.03$ \\
    
     & $\mathrm{Val}_{B}$ &  & $0.05 \pm 0.01$ &  $3.52 \pm 0.02$& $6.64 \pm 0.03$ \\
     \hline
    \end{tabular}
    }
    \caption{\label{tab:Tableseeds} Summary of the best epoch runs with random seeds (0, 1, 2) and 4 possible ensembles generated from those. Note that PEAR is the model analyzed in Sec.~\ref{sec:results} and ARA is calculated at 100~mm slices.
}
    \end{table}

\FloatBarrier

\section{Data and Code Availability}
The data and code used in this study are publicly available:
\begin{itemize}
    \item \textbf{Raw data} is generated with code available at \url{bitbucket.org/expalpha/alphasoft}.
    \item \textbf{Preprocessed data} is stored at \url{zenodo.org/records/13963779}.
    \item \textbf{Code} for data processing, model training, and generation of the presented results, can be accessed at \url{gitlab.triumf.ca/alpha-ai/rTPC-AI/-/releases/v2.0}.
\end{itemize}

\FloatBarrier

\bibliographystyle{JHEP}
\bibliography{biblio.bib}
\newpage
\end{document}